\documentstyle[aps,prl,tighten,twocolumn,psfig]{revtex}

\begin{document}
\draft
\title{Theory of Exciton Migration and Field-Induced Dissociation in 
Conjugated Polymers}
\author{M. C. J. M. Vissenberg$^{1,2}$ and M. J. M. de Jong$^1$}
\address{$^1$Philips Research Laboratories, 5656 AA Eindhoven, The
         Netherlands}
\address{$^2$Instituut-Lorentz, University of Leiden, 2300 
         RA Leiden, The Netherlands}
\date{July 8, 1996}

\maketitle

\begin{abstract}
The interplay of migration, recombination, and dissociation of 
excitons in disordered 
media is studied theoretically in the low temperature regime. 
An exact expression for the photoluminescence spectrum is obtained. 
The theory is 
applied to describe the electric field-induced photoluminescence-quenching 
experiments by Kersting {\it et al.} [Phys. Rev. Lett.
 {\bf 73}, 1440 (1994)] and
Deussen {\it et al.} [Synth. Met. {\bf 73}, 123 (1995)]
 on conjugated polymer systems. 
Good agreement with experiment is obtained using an {\em on-chain}  
dissociation  mechanism, which implies a separation of the electron-hole 
pair along the polymer chain. 
\end{abstract}

\pacs{PACS numbers: 78.66.Qn, 71.35.-y, 33.50.Hv, 78.47.+p, 
{\tt cond-mat/9607054}}

\narrowtext

Conjugated polymers (CPs) have been the subject of intensive research during 
the last decade due to their potential application in optoelectronic 
devices~\cite{Burroughes}  and the unique properties expected for 
one-dimensional (1D) systems~\cite{Heeger}. Although both 
electroluminescence (EL) 
 and photoluminescence (PL) are thought to be due to excitonic 
emission~\cite{Burroughes}, the exact nature and kinetics of 
 excitons  in these systems are still under debate.
The results of site selective fluorescence (SSF) 
experiments~\cite{Rauscher} and of time and 
spectrally resolved PL 
spectroscopy~\cite{Kersting1,Bassler} have been interpreted by 
B\"{a}ssler and co-workers in terms of a  molecular model, where    
 the polymer is treated as 
an array of sites that are  
subject to both energetic and positional disorder.
Photoabsorption  creates excitons that are localized on a site.  
These Frenkel excitons can either migrate towards sites of 
lower energy, recombine radiatively, or decay non-radiatively. 
It is believed that the observed redshift
between absorption and emission spectra is---at least for a large part---due 
to the incoherent migration of excitons before recombination.
Of specific interest regarding the EL efficiency of optoelectronic devices 
are the non-radiative decay mechanisms. 
In molecular materials, excitons can dissociate 
into free charge carriers 
through the intermediate formation of an 
indirect exciton (charge-transfer state, interchain exciton), 
i.e., a Coulombically bound electron-hole pair on two neighboring sites. 
As the formation of indirect excitons is enhanced by an applied electric field, 
the results of field-induced PL-quenching experiments on CP 
blends~\cite{Kersting2,Deussen} 
have been interpreted as experimental evidence in favor  of the 
molecular approach~\cite{Bassler}. 
However, there is an ongoing controversy about the existence of  
indirect excitons  in CPs (see, e.g., Refs.~\cite{Yan,Hayes,Blatchford}).  
Furthermore, the 1D character of the CP chains~\cite{Heeger} plays no role 
within the molecular model. 
This 1D character can give rise to excitons that are of the  Frenkel type
perpendicular to the chain direction, but of the  Mott-Wannier
type along the chain direction~\cite{Hong}, thus allowing for a {\em direct}
 1D Onsager-like dissociation process~\cite{Gailberger} 
that does not require the formation of an indirect exciton.

In this paper, we demonstrate that a {\em combination} of the molecular 
migration model and the on-chain dissociation process satisfactorily accounts 
for the field-induced PL quenching experiments of 
Kersting {\it et al.}~\cite{Kersting2} and
Deussen {\it et al.}~\cite{Deussen}.
Within our analytic theory, we derive an exact  
expression for the PL spectrum at zero temperature. 
Previously, time-dependent PL quenching~\cite{Kersting2} has been 
described theoretically by 
Arkhipov {\it et al.}~\cite{Arkhipov}. Our theory is different in two 
 respects that are essential for understanding the 
experimental results~\cite{Kersting2,Deussen}:
Firstly, unlike in Ref.~\cite{Arkhipov}, we take into account 
recombination of excitons after the first migration step.   
Secondly and most importantly, Arkhipov {\it et al.}~\cite{Arkhipov} assume 
that the formation of an indirect exciton is the primary step of  
exciton dissociation.
Our assumption of on-chain dissociation is supported by the 
experimental observation that the PL quenching in CP blends is in
fact {\em independent} of the concentration of active material, except for
very dilute blends~\cite{Deussen}. This seems to rule out the formation
of indirect excitons as the main dissociation mechanism, since it is very
sensitive to the intersite spacing.

Let us start with an outline of the theory of exciton migration in disordered 
media. 
Consider  a system with $N$ localized  states $i$, with   
 position ${\bf{R}}_i$ and  exciton energy $\varepsilon_i$.  
 For low excitation densities, the occupational probability $f_i (t)$ 
of the site $i$ at time $t$ is  
 described by a linearized master equation
\begin{equation}
\frac{ \partial}{\partial t} f_i (t) = - \sum_{j \neq i} W_{ji} f_i (t)+
 \sum_{j\neq i} W_{ij} f_j (t) - \lambda f_i (t), 
\label{eq:master}
\end{equation}
where $W_{ij}$ 
is the transition rate from site $j$ to site  $i$, and $\lambda$ is 
the loss rate due to both radiative recombination and non-radiative 
decay processes, 
 which for the moment is assumed to be constant for all sites. The transitions 
 of neutral  optical excitations are described by F\"{o}rster 
rates~\cite{Forster}, which, 
at zero temperature, have the form $W_{ij} = w_{ij}\theta (\varepsilon_j - 
\varepsilon_i)$,  with
\begin{equation}
w_{ij}  = \nu_0 {\left( \frac{R_{0}}{R_{ij}} \right)}^6   
          \label{eq:Tnulrates}
\end{equation}
and $\theta(x) = 1$ if $x > 0$, $\theta(x) = 0$ otherwise. Here, $\nu_0$  is 
the nearest neighbor jump frequency, $R_{0}$  is
the nearest neighbor distance, and    
$R_{ij} \equiv \left| {\bf{R}}_i - {\bf{R}}_j \right|$.   

Eq.~(\ref{eq:master}) can be solved  
using a Green function 
formalism~\cite{Movaghar}. 
The Green function 
 $G_{ij} (t)$ is the probability to find a particle at site $i$ at  a 
time  $t$, given that it was at site $j$ at $t = 0$, and 
 is a function of all $N$ positions and  
energies of a given configuration  $\{ {\bf{R}}_{k}, {\varepsilon}_{k} \}$. 
Assuming uncorrelated positions and energies, the configurational averages 
of the local Green function $G_{jj}(t)$ and the non-local Green function 
$G_{ij}(t)$ (with $i \neq j$) are given by
\begin{mathletters}
\begin{eqnarray}
G_1 (\varepsilon_j,t) 
      & \equiv &
      \int \left[ \prod_{k \neq j} d{\bf{R}}_{k} d{\varepsilon}_{k} 
      \frac{\rho ({\varepsilon}_{k})}{ N  } \right] G_{jj} (t)  ,  \\
G_2 ( \varepsilon_i, \varepsilon_j, {\bf{R}}_{ij}, t)  
      & \equiv &
      \int \left[ \prod_{k \neq i,j} d{\bf{R}}_{k} d{\varepsilon}_{k} 
      \frac{\rho ({\varepsilon}_{k})}{ N  } \right] G_{ij} (t)  , 
\end{eqnarray}
\end{mathletters}%
where  $\rho ({\varepsilon}_{k})$ denotes the density of localized states 
at energy ${\varepsilon}_{k}$.

The average local Green function has been  
evaluated exactly for large $N$ by F\"{o}rster~\cite{Forster},
\begin{equation}
G_1 ( {\varepsilon}_{j},t)  
= \exp{\left[ -\lambda t  - n({\varepsilon}_{j}) {\case43} 
    \pi R_0^3  \sqrt{ \pi \nu_0 t} \, \right]} , 
\label{eq:Gloc}
\end{equation}
with $n({\varepsilon}_{j}) \equiv \int_{-\infty}^{{\varepsilon}_{j}} 
d \varepsilon 
\rho ( \varepsilon )$  the density of sites with energies below 
${\varepsilon}_{j}$. 

Following Movaghar {\it et al.}~\cite{Movaghar}, 
we find for the Fourier-Laplace transform of 
$G_2 ( \varepsilon_i, \varepsilon_j, {\bf{R}}_{ij}, t)$,   
\begin{eqnarray}
{\widetilde{G}}_{2} ({\varepsilon_i}, {\varepsilon_j}, {\bf k}, p) & = &
   \theta ({\varepsilon}_{j}-{\varepsilon}_{i})
   {\widetilde{G}}_{1} (\varepsilon_i, p) \, 
   \widetilde{g}({\varepsilon}_{j}, {\bf k}, p) \nonumber \\
   && \mbox{$\times$} \exp{\left[ \int_{\varepsilon_i}^{\varepsilon_j}
   d {\varepsilon}_{l} \rho ( {\varepsilon}_{l} )
   \widetilde{g}({\varepsilon}_{l}, {\bf k}, p) \right] }  ,
   \label{eq:solGnonloc}
\end{eqnarray}
where ${\widetilde{G}}_{1} (\varepsilon_i, p)$ 
is the Laplace transform of $G_1 ( {\varepsilon}_{i},t)$ and
 $\widetilde{g}({\varepsilon}_{j}, {\bf k}, p)$ 
is the Fourier-Laplace transform of  
the average probability to jump at time $t$ from site $j$ to 
site $i$ (given that $\varepsilon_i < \varepsilon_j$).
For ${\bf k} = {\bf 0}$, one has~\cite{Movaghar}
\begin{equation}
\widetilde{g} ( {\varepsilon}_{j}, {\bf 0}, p ) = \frac{1}{n(\varepsilon_j)}
  \left[ 1 - ( p+ \lambda ) {\widetilde{G}}_{1}({\varepsilon}_{j},p) \right] .
  \label{eq:g}
\end{equation}
We note that an expansion of our result~(\ref{eq:solGnonloc}) corresponds
to the multi-step approach of Movaghar {\it et al.}~\cite{Movaghar}.

The PL intensity at  time $t$ after photoabsorption can be written as
\begin{mathletters}
\begin{equation}
L(t)= \int d{\varepsilon}_{j} L_1 ( {\varepsilon}_{j}, t) +  \int 
      d{\varepsilon}_{i} d{\varepsilon}_{j} L_2( {\varepsilon}_{i}, 
      {\varepsilon}_{j}, t) ,  \label{eq:Lum}
\end{equation}
\begin{equation}
    L_1 ( {\varepsilon}_{j}, t) \equiv \eta \lambda \Omega \rho
    ({\varepsilon}_{j}) f_j (0) G_1 ( \varepsilon_j, t)  ,
    \label{eq:energieafhlum1}
\end{equation}
\begin{eqnarray}
    L_2 ( {\varepsilon}_{i}, {\varepsilon}_{j}, t) &  \equiv & \eta \lambda
    \Omega \rho ({\varepsilon}_{i}) \rho ({\varepsilon}_{j})  f_j (0)
    \nonumber \\
    && \mbox{$\times$} \int d{\bf{R}}_{ij}  G_2 ( {\varepsilon}_{i},
    {\varepsilon}_{j}, {\bf{R}}_{ij} ,t) , \label{eq:energieafhlum2}
\end{eqnarray}
\label{eq:Lumtot}%
\end{mathletters}%
with $\eta$ the efficiency of radiative recombination and $\Omega$ the 
system volume. 
The PL intensity 
$L_1 ( {\varepsilon}_{j}, t)$ is due to  
 excitons that 
recombine at their initial site with energy ${\varepsilon}_{j}$ and 
$L_2 ( {\varepsilon}_{i}, {\varepsilon}_{j}, t)$ is from  
excitons that recombine on a site with energy ${\varepsilon}_{i}$ 
after migration from the initial site with energy ${\varepsilon}_{j}$.  
Note that the latter contribution  has not been taken into account 
in Ref.~\cite{Arkhipov}.

Eqs.~(\ref{eq:Gloc})--(\ref{eq:Lumtot}) allow us to study 
PL spectra for arbitrary initial condition $f_j (0)$  and  
density of localized states $\rho ( {\varepsilon})$. 
In most cases, the system is excited  either by a broad-band lightsource,  
$f_j(0) = f^0$, or by a narrow-width laser pulse with 
energy $\varepsilon_0$,    
$f_j(0) = f^0 \delta ( \varepsilon_j - \varepsilon_0 )$. 
For example, SSF experiments~\cite{Rauscher} can be described by varying 
the excitation energy $\varepsilon_0$. 
Time-integrated PL spectra can be evaluated directly by substituting $p=0$ 
in the Laplace transforms of Eqs.~(\ref{eq:Lumtot}), whereas time resolution 
can be obtained using a numerical Laplace inversion routine. 

In the following, our theory of exciton migration is used to interpret 
 field-induced PL-quenching experiments by Deussen 
{\it et al.}~\cite{Deussen} and Kersting {\it et al.}~\cite{Kersting2}. 
Both study the optical response of thin-film light-emitting diode 
structures made from blends of poly(phenyl-{\it p}-phenylene vinylene) (PPPV)
and polycarbonate (PC), initially excited by a laser pulse 
with energy $\varepsilon_0$. 
We assume---as they did---that the polymers are perfectly mixed. 
We further expect the zero-temperature approach to be valid as the 
experiments are performed at 70--80 K,  
whereas typical energy differences are of order 0.1 eV.

\begin{figure}[h]
\begin{center}
\mbox{\psfig{file=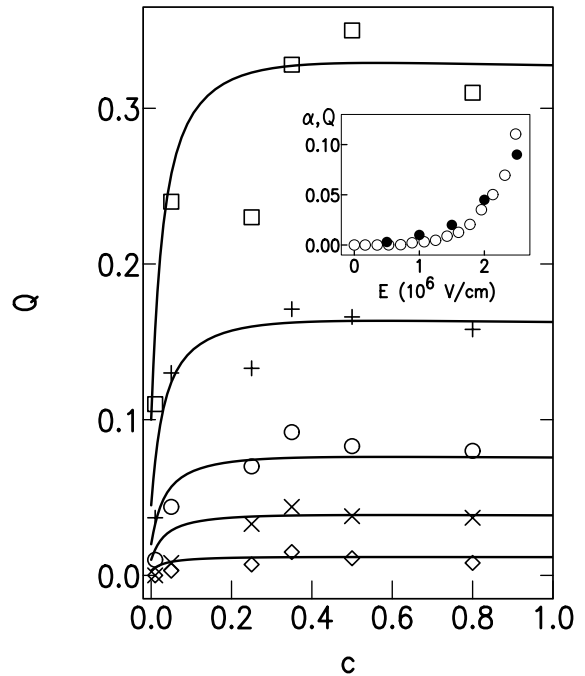,angle=0,width=7.4cm}}
\end{center}
\caption{Photoluminescence quenching $Q$ as a function of concentration $c$
 for an electric field $E=0.5, 1.0, 1.5, 2.0,$  and $2.5$ MV/cm 
(bottom to top).
The parameters used are $n(\varepsilon_0)/n = 0.9$,
$\lambda^{'} = 10^{-4}/c^2$, and $\lambda_d^{'} = 0.1/c^2$. The values
of $\alpha$ are given in the inset (dots), together with
 $Q$ for $c = 0.01$ (circles).
Experimental data are taken from Ref.~\protect\cite{Deussen}. }
\label{fig1}
\end{figure}

\begin{figure}[h]
\begin{center}
\mbox{\psfig{file=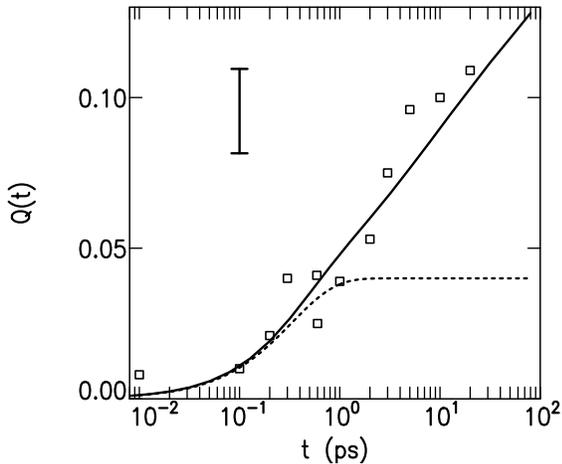,angle=0,width=7.4cm}}
\end{center}
\caption{Transient photoluminescence quenching $Q(t)$ as a
function of time $t$ (solid line).
The same parameters
are used as in Fig. 1, with $c = 0.2$ and $\alpha = 0.04$.
The dashed line denotes the quenching when
the luminescence of migrated excitons, $L_2(\varepsilon_i, \varepsilon_0, t)$,
is neglected.
Experimental data are taken from Ref.~\protect\cite{Kersting2}.
An indication of the experimental error is given in the upper left corner. }
\label{fig2}
\end{figure}

 In the presence of an applied 
electric field, the PL intensity of these systems, $L^q(t)$, is reduced. 
In Ref.~\cite{Deussen}, the relative decrease of the {\em time-integrated} 
PL intensity, 
\begin{equation}
Q = \frac{{\widetilde{L}}(p = 0) -
    {\widetilde{L}}^q(p = 0)}{{\widetilde{L}}(p = 0)} ,
\label{eq:quenching}
\end{equation}
is studied as a function of the excitation energy $\varepsilon_0$, 
the relative concentration $c \in [0,1]$ of active material (PPPV), 
and the strength of the 
electric field $E$. In Fig.~\ref{fig1}, the concentration dependence of $Q$ 
is given for various field strengths.   
In  Ref.~\cite{Kersting2}, the {\em transient} PL quenching,  
\begin{equation}
Q(t) = \frac{L(t) - L^q(t)}{L(t)} ,
\label{eq:transientQ}
\end{equation}
is studied for a system with fixed $\varepsilon_0$, $c$, and $E$. Their 
results are given in Fig.~\ref{fig2}.   
 
In the absence of an electric field, the PL intensity is described by
\begin{eqnarray}
{\widetilde{L}} (p) &=& \eta \lambda \Omega f^0  \rho ({\varepsilon}_0) \left\{
    {\widetilde{G}}_{1} ( {\varepsilon}_{0},p) + 
    {\widetilde{g}} ({\varepsilon}_{0}, {\bf 0}, p ) 
     \int_{-\infty}^{\varepsilon_0} d \varepsilon_i \rho ({\varepsilon}_i)
     \right. \nonumber \\
    && \mbox{$\times$} \left.  {\widetilde{G}}_{1} ( {\varepsilon}_{i}, p)  
    \exp{\left[ \int_{{\varepsilon}_{i}}^{ {\varepsilon}_{0}} 
    d {\varepsilon}_{l} \rho ({\varepsilon}_{l}) 
    {\widetilde{g}} ({\varepsilon}_{l}, {\bf 0}, p ) \right] } \right\} .
  \label{eq:laser}
\end{eqnarray}
The energy integration can be evaluated exactly for arbitrary 
$\rho (\varepsilon)$, resulting in an  
exponential time decay as found in PL experiments on  CP blends and 
solutions (see, e.g., Refs.~\cite{Yan,Blatchford}). 
From time-resolved PL 
experiments~\cite{Kersting1,Yan,Hayes,Blatchford}, we estimate  
  $\lambda \approx {(300 \: {\rm ps})}^{-1}$ and  $\nu_0 \approx 10^{13}$ Hz. 
We take $n \case43 \pi R_0^3 = c$, with 
$n \equiv n(\varepsilon{\rightarrow}\infty) = N/\Omega$. 

To describe the quenched PL intensity in the presence of  
an electric field, our key assumption is an on-chain dissociation of excitons.
One may expect that the dissociation of an exciton  on a given site depends on 
the strength of the electric field, 
the orientation of the CP chain segment with respect to the electric field, 
and the exciton binding energy.   
We model this as follows: dissociation can only occur on a fraction 
$\alpha$ of the sites with a loss rate $\lambda_d$, so that the total 
loss rate is $\lambda + \lambda_d$; 
on the other sites the total loss rate is just $\lambda$. 
The specific on-chain character of the dissociation process is reflected in
the concentration independence of the parameters $\lambda_d$ and $\alpha$.
In the following, we will just fit these parameters to the experiments,
leaving a detailed physical interpretation to future work.
We do not take into account the second-order Stark effect, which leads to 
an additional redshift in the PL spectrum and a reduction of $\eta$. 
  
When dissociation sites are present, the average local Green function reads  
\begin{equation}
{\widetilde{G}}_{1}^q ( {\varepsilon}_{j},p) = (1-\alpha) {\widetilde{G}}_{1} 
( {\varepsilon}_{j},p) + \alpha {\widetilde{G}}_{1}^d ( {\varepsilon}_{j},p) , 
\label{eq:qG}
\end{equation}
where ${\widetilde{G}}_{1}^d ( {\varepsilon}_{j},p)$ 
 is the Laplace transform of Eq.~(\ref{eq:Gloc}) 
  with $\lambda$ replaced by $\lambda + \lambda_d$.  
For the average probability to  jump from  site $j$ one has 
\begin{eqnarray}
{\widetilde{g}}^q ( {\varepsilon}_{j}, {\bf 0}, p ) &  = &  
 \frac{1}{n(\varepsilon_j)}
 \left[ 1 - (1-\alpha) ( p+ \lambda ) {\widetilde{G}}_{1}({\varepsilon}_{j},p)
 \right. \nonumber \\
 && \left. \mbox{$-$}  \alpha ( p+ \lambda + \lambda_d )
  {\widetilde{G}}_{1}^d ({\varepsilon}_{j},p) \right] . 
   \label{eq:qg}
\end{eqnarray}
The quenched PL intensity, 
${\widetilde{L}}^q (p)$,  
follows from  substitution of  
Eqs.~(\ref{eq:qG}) and~(\ref{eq:qg}) into Eq.~(\ref{eq:laser}). 

Although many microscopic parameters have been introduced, we 
find from Eq.~(\ref{eq:quenching}) 
that $Q$ is a function of only four independent parameters: 
the fraction of sites with energies below the 
excitation energy $n(\varepsilon_0)/n$;  the fraction of 
dissociation sites $\alpha$; 
the relative recombination and dissociation rates $\lambda^{'}$ 
and $\lambda_d^{'}$,  
 given by 
\begin{equation}
\frac{ \lambda}{\lambda^{'}} = \frac{ \lambda_d}{\lambda_d^{'}} =  
 \pi \nu_0 {( n \case43 \pi R_0^3 )}^2  . 
\end{equation}
As an important consequence, our expression for the PL quenching is 
{\em independent} of the form of the density of states 
 $\rho (\varepsilon)$.  Taking experimental parameters, we estimate 
$\lambda^{'} = 10^{-4}/c^2$.  

Conforming to the results of Ref.~\cite{Deussen}, we find that $Q$ is 
 approximately  constant  as a function of excitation energy  
[or $n(\varepsilon_0)/n$] except for very low energies.  
Consequently, the PL quenching is  
 only weakly dependent on the initial condition $f_j(0)$.

The concentration dependence of $Q$, as given in Fig.~\ref{fig1}, can be 
understood as follows: 
At very low $c$, the excitons are immobile ($\lambda^{'} \rightarrow \infty$) 
and $Q$  is given by the quenching efficiency of an isolated site, 
\begin{equation}
\lim_{c \rightarrow 0} Q = \alpha \frac{\lambda_d}{\lambda + \lambda_d} .
\label{eq:alpha}
\end{equation}
With increasing $c$, the excitons visit more sites during their 
lifetime yielding a  higher probability of an encounter
with a dissociation site, so that $Q$ increases. 
As $c$ further increases, migration from dissociation sites before 
dissociation takes place results in a saturation of $Q$. 
This competition between dissociation and migration depends on 
$\lambda_d^{'}$.

In Fig.~\ref{fig1}, the experimental values of $Q$ as a function of $c$ 
for various $E$ are compared with our theory. 
Given the minimal assumptions put in our model, the agreement is quite good. 
The values of $\lambda_d^{'}$ and $\alpha$ are determined from 
the onset of saturation at $c \approx 0.3$ and the saturation values of $Q$ 
for different $E$. 
We find good agreement using
$\lambda_d^{'} = 0.1/c^2$ for all values of $E$, which
corresponds to $\lambda_d \approx {(300 \: {\rm fs})}^{-1}$.  
The values of $\alpha$ for each $E$ are depicted in the inset of 
Fig.~\ref{fig1}. 
According to Eq.~(\ref{eq:alpha}), these values should match the 
experimental values of $Q$ at $c = 0.01$ (see the inset of Fig.~\ref{fig1}). 
We note that further improvement may be achieved taking into account, 
e.g., an $E$-dependence of $\eta$ or a sample-dependent trap density. 
We stress that  recombination of migrating excitons, i.e. 
$L_2(\varepsilon_i, \varepsilon_0, t)$, is required to describe the 
increase of $Q$ at low $c$ and that    
 the concentration-independence of the dissociation 
process is essential to obtain the saturation of $Q$ at high $c$.  
 
Let us make the comparison with the experimental results of Kersting 
{\it et al.}~\cite{Kersting2} 
 on   a PPPV/PC sample with $c = 0.2$.   
We simply use the same set of parameters as above. 
 The time-integrated PL quenching $Q = 0.135$ yields $\alpha = 0.04$. 
The transient PL quenching $Q(t)$ 
[Eq.~(\ref{eq:transientQ})] is compared with experiment in Fig.~\ref{fig2}. 
Given the fact that we have not used any fitting parameters, the 
agreement is remarkable.  
For comparison, we have also plotted  the quenching when 
 recombination {\em after} migration, i.e.  
 $L_2(\varepsilon_i, \varepsilon_0, t)$, is not taken into account.  
We see that, after the first picosecond, the increase of $Q(t)$ is entirely 
due to the migration of excitons, which, for long times, leads to 
$Q(t\rightarrow\infty)=1$~\cite{remark}. 
This is in stark contrast with the model proposed in Ref.~\cite{Arkhipov}, 
where $Q(t)$ saturates towards the value of $Q$, but it is in 
agreement with the experimental findings of 
Kersting {\it et al.}~\cite{Kersting2},  who report a further  PL  quenching  
on a timescale of several hundred picoseconds.

In summary, we have presented an  analytic zero-temperature theory 
of exciton migration and PL in a disordered medium which 
can be used to interpret  
a variety of spectroscopic experiments. 
The theory has been applied to describe  
 field-induced PL-quenching experiments in conjugated polymer blends. 
We find that the effect is governed by a subtle balance of 
recombination, migration, and {\em on-chain}  dissociation and that 
close agreement with experiment is obtained for realistic values 
of the parameters involved. 
 
We acknowledge useful discussions with C. W. J. Beenakker and 
G. H. L. Brocks. This work was supported by the  
Dutch Science Foundation NWO/FOM.


\begin{references}
\bibitem{Burroughes}J. H. Burroughes, D. D. C. Bradley, A. R. Brown, 
R. N. Marks, K. MacKay, R. H. Friend, P. L. Burn, and A. B. Holmes, 
Nature (London) {\bf 347}, 539 (1990). 
\bibitem{Heeger}A. J. Heeger, S. Kivelson, J. R. Schrieffer, and W. P. Su, 
Rev. Mod. Phys. {\bf 60}, 781 (1988).
\bibitem{Rauscher}U. Rauscher, H. B\"{a}ssler, D. D. C. Bradley, and 
M. Hennecke, Phys. Rev. B {\bf 42}, 9830 (1990); 
\bibitem{Kersting1}R. Kersting, U. Lemmer, R. F. Mahrt, K. Leo, H. Kurz, 
H. B\"{a}ssler, and E. O. G\"{o}bel, Phys. Rev. Lett. {\bf 70}, 3820 (1993).
\bibitem{Bassler}H. B\"{a}ssler, V. Brandl, M. Deussen, E. O. G\"{o}bel, 
R. Kersting, H. Kurz, U. Lemmer, R. F. Mahrt, and A. Ochse, Pure \& Appl. 
Chem. {\bf 67}, 377 (1995).
\bibitem{Kersting2}R. Kersting, U. Lemmer, M. Deussen, H. J. Bakker, 
R. F. Mahrt, H. Kurz, V. I. Arkhipov, H. B\"{a}ssler, and E. O. G\"{o}bel, 
Phys. Rev. Lett. {\bf 73}, 1440 (1994).
\bibitem{Deussen}M. Deussen, M. Scheidler, and H. B\"{a}ssler, Synth. Met. 
{\bf 73}, 123 (1995).
\bibitem{Yan}M. Yan, L. J. Rothberg, E. W. Kwock, and T. M. Miller, 
Phys. Rev. Lett. {\bf 75}, 1992 (1995).
\bibitem{Hayes}G. R. Hayes, I. D. W. Samuel, and R. T. Philips, 
Phys. Rev. B {\bf 52}, R11569 (1995).
\bibitem{Blatchford}J. W. Blatchford, S. W. Jessen, L. B. Lin, J. J. Lih, 
T. L. Gustafson, A. J. Epstein, D. K. Fu, M. J. Marsella, T. M. Swager, 
A. G. MacDiarmid, S. Yamaguchi, and H. Hamaguchi, Phys. Rev. Lett. {\bf 76}, 
1513 (1996).
\bibitem{Hong}H. X. Wang and S. Mukamel, Chem. Phys. Lett. {\bf 192}, 
417 (1992); \'{A}. Horv\'{a}th, H. B\"{a}ssler, and G. Weiser, 
Phys. Stat. Sol. (b) {\bf 173}, 755 (1992); D. Beljonne, Z. Shuai, 
R. H. Friend, and J. L. Br\'{e}das, J. Chem. Phys. {\bf 102}, 2042 (1995).
\bibitem{Gailberger}M. Gailberger and H. B\"{a}ssler, Phys. Rev. B {\bf 44}, 
8643 (1991).
\bibitem{Arkhipov}V. I. Arkhipov, H. B\"{a}ssler, M. Deussen, E. O. G\"{o}bel, 
R. Kersting, H. Kurz, U. Lemmer, and R. F. Mahrt, Phys. Rev. B {\bf 52}, 
4932 (1995).
\bibitem{Forster}Th. F\"{o}rster, Z. Naturforschg. {\bf 4a}, 321 (1949).
\bibitem{Movaghar}B. Movaghar, B. Ries, and M. Gr\"{u}newald, Phys. Rev. B 
{\bf 34}, 5574 (1986); M. Gr\"{u}newald and B. Movaghar, 
J. Phys. Condens. Matter {\bf 1}, 2521 (1989).
\bibitem{remark}For the intermediate time regime, Fig. 2 suggests that 
$Q(t) \propto \ln{(t)}$. Indeed, one can derive that, up to linear order in 
$\alpha$, $Q(t) = \alpha \left[ 1+ \case12 \ln{\left( 1+\pi \nu_0 {\left[
 c n(\varepsilon_0) / n \right]}^2 t \right)} \right]$.
\end{references}
\end{document}